\documentclass[12pt]{article}
\title{Quantized Feedback Control Software Synthesis from System Level Formal Specifications for Buck DC/DC Converters}

\author{Federico Mari, Igor Melatti, Ivano Salvo, Enrico Tronci\\
\small \itshape Department of Computer Science\\
\small \itshape Sapienza University of Rome\\
\small \itshape via Salaria 113, 00198 Rome\\
\small email: \{mari,melatti,salvo,tronci\}@di.uniroma1.it}

\usepackage{color}
\usepackage{balance}
\usepackage{latexsym}
\usepackage{graphicx}
\usepackage{epsfig}
\usepackage[hang]{subfigure}
\usepackage{algorithm}
\usepackage[noend]{algorithmic}
\usepackage[T1]{fontenc}
\usepackage{epic}
\usepackage{eepic}
\usepackage{amsfonts}
\usepackage{amssymb}
\usepackage{stmaryrd}
\usepackage{amsthm}
\usepackage[all]{xy}
\usepackage{booktabs}
\usepackage{multirow}
\usepackage{rotating}
\usepackage{multicol}

\newtheorem{theorem}{Theorem}
\newtheorem{proposition}[theorem]{Proposition}

\newtheorem{definition}{Definition}

\algsetup{linenodelimiter=.} 
\algsetup{indent=1em}

\definecolor{light-gray}{gray}{0.90}
\newcounter{theorem-backup}

\newcommand{\R}{\mathbb{R}}

\newcommand{\B}{\mathbb{B}}

\usepackage{listings}
\lstdefinelanguage{PseudoC}[ISO]{C++} { 
morekeywords={foreach, and, not, or, is, FIFO_Queue, HashTable, FILE, Cache}, 
morekeywords=[2]{HashInsert, Enqueue, Dequeue, next}, 
mathescape=true,
alsoletter={'},
deletestring=[b]'
}

\lstdefinelanguage{PseudoCWithNumbers}[ISO]{C++} { 
morekeywords={foreach, and, not, or, is, FIFO_Queue, HashTable, FILE, Cache}, 
morekeywords=[2]{HashInsert, Enqueue, Dequeue, next}, 
mathescape=true,
alsoletter={'},
deletestring=[b]'
}

\lstdefinelanguage{Murphi}[]{Pascal} { 
morekeywords={ruleset, rule, invariant, startstate, return, endif, endfor, endswitch, forall, endforall, exists, endexists}, 
mathescape=true, 
morestring=[b]",
morestring=[b]', 
morecomment=[s]{/*}{*/} ,
morecomment=[l]{--}
}

\lstdefinelanguage{PRISM}[ISO]{C++} { 
morekeywords={probabilistic, stochastic, const, rate, module, endmodule, init, P, U},
mathescape=true, 
alsoletter={'},
deletestring=[b]'
}

\lstdefinelanguage{Yacc}[ISO]{C++} { 
morekeywords={token, left}, 
mathescape=false,
alsoletter={'},
deletestring=[b]'
}

\lstdefinestyle{PseudoC}{
language=PseudoC,
basicstyle=\ttfamily,%\small, 
tabsize=1,
showlines=false,
emptylines=*1,
breaklines=true,
breakindent=5pt,
keywordstyle=\rmfamily\bfseries,%\normalsize, 
keywordstyle=[2]\rmfamily,%\normalsize, 
commentstyle=\itshape, 
columns=fixed,
showspaces=false, 
showstringspaces=false, 
showtabs=false, 
escapechar=\%
}

\lstdefinestyle{PseudoCWithNumbers}{
language=PseudoC,
basicstyle=\ttfamily,%\small, 
tabsize=1,
showlines=false,
emptylines=*1,
breaklines=true,
breakindent=5pt,
keywordstyle=\rmfamily\bfseries,%\normalsize, 
keywordstyle=[2]\rmfamily,%\normalsize, 
commentstyle=\itshape, 
columns=fixed,
showspaces=false, 
showstringspaces=false, 
showtabs=false, 
escapechar=\%,
numbersep=5pt,framexleftmargin=15pt,numbers=left,
}

\lstdefinestyle{Murphi}{
language=Murphi,
basicstyle=\ttfamily,%\small, 
tabsize=1,
showlines=false,
emptylines=*1,
breaklines=true,
breakindent=5pt,
keywordstyle=\rmfamily\bfseries,%\normalsize, 
commentstyle=\itshape, 
columns=fixed,
showspaces=false, 
showstringspaces=false, 
showtabs=false,
escapechar=\%}

\lstdefinestyle{PRISM}{
language=PRISM,
basicstyle=\ttfamily,%\small, 
tabsize=1,
showlines=false,
emptylines=*1,
breaklines=true,
breakindent=5pt,
keywordstyle=\rmfamily\bfseries,%\normalsize, 
commentstyle=\itshape, 
columns=fixed,
showspaces=false, 
showstringspaces=false, 
showtabs=false,
escapechar=\%}

\lstdefinestyle{Yacc}{
language=PseudoC,
basicstyle=\ttfamily,%\small, 
tabsize=1,
showlines=false,
emptylines=*1,
breaklines=true,
breakindent=5pt,
keywordstyle=\rmfamily\bfseries,%\normalsize, 
keywordstyle=[2]\rmfamily,%\normalsize, 
commentstyle=\itshape, 
columns=fixed,
showspaces=false, 
showstringspaces=false, 
showtabs=false, 
}

\newcommand{\eqref}[1]{(\ref{#1})}

\newcommand{\qks}{\mbox{QKS}}

\begin{document}

\maketitle

\begin{abstract}

Many \emph{Embedded Systems} are indeed \emph{Software Based Control Systems}
(SBCSs), that is control systems whose controller consists of \emph{control
software} running on a  microcontroller device.  This motivates investigation on
\emph{Formal Model Based Design}  approaches for automatic synthesis of SBCS
control software. 
%
%We present an algorithm, along with a tool \qks \ implementing it, that 
%from a formal model (as a \emph{Discrete Time Linear Hybrid System}, DTLHS)
%of the controlled system (\emph{plant}), 
%\emph{implementation specifications} 
%(that is, number of bits in the \emph{Analog-to-Digital}, AD, conversion) and
%\emph{System Level Formal Specifications}
%(that is, safety and liveness requirements for the \emph{closed loop system})
%returns correct-by-construction control software that has a
%\emph{Worst Case Execution Time} (WCET) linear in the number of AD bits
%and meets the given specifications. 
%
%We show feasibility
%of our approach by presenting experimental results on using it 
%to synthesize control software for a buck DC-DC converter, 
%a widely used mixed-mode analog circuit.
%
In previous works we presented an algorithm, along with a tool \qks \
implementing it, that  from a formal model (as a \emph{Discrete Time Linear
Hybrid System}, DTLHS) of the controlled system (\emph{plant}), 
\emph{implementation specifications}  (that is, number of bits in the
\emph{Analog-to-Digital}, AD, conversion) and \emph{System Level Formal
Specifications} (that is, safety and liveness requirements for the \emph{closed
loop system}) returns correct-by-construction control software that has a
\emph{Worst Case Execution Time} (WCET) linear in the number of AD bits and
meets the given specifications. 
In this technical report we present full experimental results on using it  to
synthesize control software for two versions of buck DC-DC converters 
(single-input and multi-input),  a widely used mixed-mode analog circuit.

\end{abstract}

\section{Introduction}\label{intro.tex}

Many \emph{Embedded Systems} are indeed \emph{Software Based
Control Systems} (SBCSs). An SBCS  consists of two main subsystems: the
\emph{controller} and the \emph{plant}. Typically, the plant is a physical
system consisting, for example, of mechanical or electrical devices whereas the
controller  consists of \emph{control software} running on a microcontroller. In
an endless loop,  the controller reads \emph{sensor} outputs from the plant and
sends commands to plant \emph{actuators} in order to guarantee that the 
\emph{closed loop system} (that is, the system consisting of both plant and
controller) meets given \emph{safety} and \emph{liveness} specifications
(\emph{System Level Formal Specifications}).

Software generation from models and formal specifications forms the core of
\emph{Model Based Design} of embedded software \cite{Henzinger-Sifakis-fm06}.
This approach is particularly interesting for SBCSs since in such a case system
level (formal) specifications are much easier to define than the control
software behavior itself. 

Fig. \ref{control-loop-figure.tex} shows the  typical control loop skeleton for
an SBCS. Measures from plant \emph{sensors} go through an AD
(\emph{analog-to-digital}) conversion (\emph{quantization}) before being
processed (line \ref{quantized-feedback}) and commands from the control software
go through a  DA (\emph{digital-to-analog}) conversion before being sent to
plant \emph{actuators} (line \ref{DA-conversion}). Basically, the control
software design problem for SBCSs consists in designing software implementing
functions \texttt{Control\_Law} and \texttt{Controllable\_Region} computing,
respectively, the command to be sent to the plant (line \ref{control-law}) and
the set of states on which the \texttt{Control\_Law} function works correctly
(\emph{Fault Detection} in line \ref{fault-detection}).

%\marginpar{Fig.~\ref{control-loop-figure.tex} o cappucciamo tutte le cose discrete (quindi anche $u$), 
%o niente, visto che qui ancora non si accenna a nulla}

%\begin{algorithm}
\begin{figure}
\line(1, 0){350}
%\medskip
%  \framebox[1.0\hsize][c]{

    \begin{minipage}{1.0\hsize}
%      \begin{center}
  \begin{algorithmic}[1]
    \STATE {\bf Every} $T$ seconds (\emph{sampling time)} {\bf do} 
                           \label{sampling-and-hold}
    \STATE \hspace*{2ex} {\bf Read} AD conversion $\hat{x}$ of plant sensor outputs $x$
                         \label{quantized-feedback}
    \STATE \hspace*{2ex} {\bf If} ($\hat{x}$ is not in the \texttt{Controllable\_Region})  
                         \label{fault-detection}
    \STATE \hspace*{4ex} {\bf Then} {\em // Exception (Fault Detected):} 
    \STATE \hspace*{6ex}      Start Fault Isolation and Recovery (\texttt{FDIR})
    \STATE \hspace*{4ex} {\bf Else} {\em // Nominal case:}
    \STATE \hspace*{6ex}  Compute (\texttt{Control\_Law}) command $\hat{u}$ from $\hat{x}$
                           \label{control-law}
    \STATE \hspace*{6ex} {\bf Send} DA conversion $u$ of $\hat{u}$ to plant actuators
                         \label{DA-conversion}
  \end{algorithmic}
%\end{center}
\end{minipage}

%\smallskip
\line(1, 0){350}

%}   %  end framebox
  \caption{A typical control loop skeleton}
  \label{control-loop-figure.tex}
\end{figure}

In~\cite{qks-cav2010} we presented an algorithm and a tool \qks \ that from the
plant model (as a hybrid system),  from formal specifications for the closed
loop system behaviour  (\emph{System Level Formal Specifications})  and from
\emph{implementation specifications}  (that is, number of bits used in the
quantization process) can generate correct-by-construction control software
satisfying the given specifications. 

In this technical report we present full experimental results on using it  to
synthesize control software for two versions of buck DC-DC converters 
(single-input and multi-input), a widely used mixed-mode analog circuit.

\section{Background}\label{basic.tex}

We denote with $[n]$ an initial segment $\{1,\ldots, n\}$ of the natural
numbers.  We denote with $X$ = $[x_1, \ldots, x_n]$ a finite sequence (list) of
variables. By abuse of language we may regard sequences as sets and we use
$\cup$ to denote list concatenation. Each variable $x$ ranges on a known
(bounded or unbounded) interval ${\cal D}_x$ either of the reals or of the
integers (discrete variables).  We denote with ${\cal D}_X$ the set $\prod_{x\in
X} {\cal D}_x$. To clarify that a variable $x$ is {\em continuous}  (i.e. real
valued) we may write $x^{r}$. Similarly,  to clarify that a variable $x$ is {\em
discrete}  (i.e. integer valued) we may write $x^{d}$.  Boolean variables are
discrete variables ranging on the set $\B$ = \{0, 1\}.  We may write $x^{b}$ to
denote a boolean variable. Analogously $X^{r}$ ($X^{d}$, $X^{b}$) denotes the
sequence of real (integer, boolean) variables in $X$. Finally, if $x$ is a
boolean variable  we write $\bar{x}$ for $(1 - x)$.

\subsection{Predicates}\label{subsection:predicates}
 
A {\em linear expression} over a list of variables $X$ is a linear combination
of variables in $X$ with real coefficients.  A {\em linear constraint} over $X$
(or simply a {\em constraint})  is an expression of the form $L(X) \leq b$,
where $L(X)$ is a linear expression over $X$  and $b$ is a real constant.

{\em Predicates} are inductively defined as follows. A constraint $C(X)$ over a
list of variables $X$ is a predicate over  $X$.  If $A(X)$ and $B(X)$ are
predicates over $X$, then $(A(X) \land B(X))$ and $(A(X) \lor B(X))$ are
predicates over X.  Parentheses may be omitted, assuming usual associativity and
precedence rules of logical operators. A {\em conjunctive predicate} is a
conjunction of constraints. For linear constraints we write: $L(X) \geq b$ for
$-L(X) \leq -b$, $L(X) = b$ for (($L(X) \leq b$) $\wedge$ ($-L(X) \leq -b$)) and
$a \leq x \leq b$ for $x \geq a \land x \leq b$, being $x \in X$.

A {\em valuation} over a list of variables $X$ is a function $v$ that maps each
variable $x \in X$ to a value $v(x)$ in ${\cal D}_x$. We denote with $X^\ast\in
{\cal D}_X$ the sequence of values  $[v(x_1),\ldots,v(x_n)]$. By abuse of
language,  we call valuation also the sequence of values $X^\ast$. A
\emph{satisfying assignment} to a predicate $P$ over $X$ is a valuation $X^{*}$
such that $P(X^{*})$ holds.  Abusing notation, we may denote with $P$ the set of
satisfying assignments to the predicate  $P(X)$. Two predicates $P$ and $Q$ over
$X$ are {\em equivalent}, notation $P\equiv Q$, if they have the same set of 
satisfying assignments.

A variable $x\in X$  is said to be {\em bounded} in $P$ if  there exist $a$, $b
\in {\cal D}_x$ such that $P(X)$ implies $a \leq x \leq b$. A predicate $P$ is
bounded if all its variables are bounded.

Given a constraint $C(X)$ and a fresh boolean variable ({\em guard}) $y \not\in
X$, the {\em guarded constraint} $y \to C(X)$ (if $y$ then $C(X)$) denotes the
predicate $((y = 0) \lor C(X))$. Similarly, we use $\bar{y} \to C(X)$ (if not
$y$ then $C(X)$) to denote the predicate $((y = 1) \lor C(X))$. A {\em guarded
predicate} is a conjunction of  either constraints or guarded constraints. 

When a guarded predicate is bounded, it can be easily transformed into a
conjunctive predicate, as stated by the following proposition. 

\begin{proposition}\label{predconjunctive.prop}

For each bounded guarded predicate $P(X)$,  there exists an equivalent bounded
conjunctive predicate $Q(X)$. 

\end{proposition}

%\begin{proof}
%
%Predicate $Q(X)$ is obtained from the guarded predicate $P(X)$ by replacing
%each  guarded constraint $\varphi$ in $P(X)$ with an equivalent linear
%constraint $\varphi^{*}$. We construct such a linear constraint $\varphi^{*}$ as
%follows. Let $x \in X$. Since $P(X)$ is bounded there exist $m_x$, $M_x  \in
%{\cal D}_x$ such that $P(X)$ implies $m_x \leq x \leq M_x$. Let $a$ be a real
%number and $x \in X$.  We write  $\sup(a x)$ [$\inf(a x)$] for  $a M_x$ [$a
%m_x$] when $a \geq 0$ and for  $a m_x$ [$a M_x$] when $a < 0$.  Let $L(X)$ =
%$\sum_{i=1}^{n}a_ix_i$ be a linear expression.  We write $\sup(L(X))$ for
%$\sum_{i=1}^{n} \sup(a_i x_i)$ and  $\inf(L(X))$ for $\sum_{i=1}^{n} \inf(a_i
%x_i)$. Let $\varphi$  be $z \rightarrow (L(X) \leq b)$.  We pick $\varphi^{*}$
%to be the linear constraint $(\sup(L(X)) - b)z + L(X) \leq \sup(L(X))$.   If
%$z=0$ we have $\varphi \equiv \varphi^{*}$ since $\varphi$ holds trivially and
%$\varphi^{*}$ reduces to $L(X) \leq \sup(L(X))$ that holds by construction.  If
%$z=1$ both $\varphi$ and $\varphi^{*}$ reduce to  $L(X) \leq b$.  Along the same
%line of reasoning, if $\varphi$  has form $\bar{z} \rightarrow (L(X) \leq b)$ 
%we pick $\varphi^{*}$ to be $(b - \sup(L(X)))z + L(X) \leq b$.
%
%\end{proof}

\section{Discrete Time Linear Hybrid Systems} \label{dths.tex}

In this section we introduce our class of  {\em Discrete Time Linear Hybrid
Systems} (DTLHS for short).
%,  together with the DTLHS representing the buck DC-DC
%converter  on which our experiments will focus.

\begin{definition}\label{dths.def}

A {\em Discrete Time Linear Hybrid System} is a tuple ${\cal H} = (X,$ $U,$ $Y,$
$N)$ where:

\begin{itemize}

\item  $X$ = $X^{r} \cup X^{d} \cup X^{b}$   is a finite sequence of real
($X^{r}$), discrete ($X^{d}$) and boolean ($X^{b}$) {\em present state}
variables.   We denote with $X'$ the sequence of  {\em next state} variables
obtained  by decorating with $'$ all variables in $X$.

\item  $U$ = $U^{r} \cup U^{d} \cup U^{b}$ is a finite sequence of  \emph{input}
variables.

\item  $Y$ = $Y^{r} \cup Y^{d} \cup Y^{b}$ is a finite sequence of
\emph{auxiliary} variables.  Auxiliary variables are typically used to model
\emph{modes} (e.g., from switching elements such as diodes)  or
\emph{uncontrollable inputs} (e.g., disturbances).

\item  $N(X, U, Y, X')$ is a conjunctive predicate  over $X \cup U \cup Y \cup
X'$ defining the  {\em transition relation} (\emph{next state}) of the system.

\end{itemize}

A DTLHS is {\em bounded} if predicate $N$ is bounded.

\end{definition}

By Prop.~\ref{predconjunctive.prop}, any bounded guarded predicate can  be
transformed into a conjunctive predicate. For the sake of readability, we will 
use bounded guarded predicates to describe the transition relation of  bounded
DTLHSs.  Note that DTLHSs can effectively model linear algebraic constraints
involving both continuous as well as discrete variables. Therefore many embedded
control systems may be modeled as DTLHSs. 
%In the following definition, we give
%the semantics of  DTLHSs in terms of LTSs. 

%\begin{example}\label{ex:robust-dths}
%
%Let ${\cal H}=(\{x\},\{u\},\varnothing,N)$ with $x$ continuous variable, $u$
%boolean variable, and $N(x, u, x')$ = $[\overline{u} \rightarrow x' = \alpha x] 
%\land [u \rightarrow x' = \beta x]$ with $\alpha = \frac{1}{2}$ and $\beta =
%\frac{3}{2}$. Adding nondeterminism to ${\cal H}$ allows us to address synthesis
%of \emph{robust controllers}, thus guaranteeing  that the closed loop system 
%meets its requirements notwithstanding (small) variations in the plant
%parameters. For example, variations in the parameter $\alpha$ can be modelled
%with a tolerance  $\rho \in [0, 1]$ for $\alpha$.  This replaces $N$ with:
%$N^{\rho}$ =  $[\overline{u} \rightarrow x' \leq (1 + \rho) \alpha x]$  $\land$
%$[\overline{u} \rightarrow x' \geq (1 - \rho) \alpha x]$  $\land$ $[u
%\rightarrow x' = \beta x]$. A controller for the DTLHS ${\cal H}^{\rho}$ = 
%$(\{x\},\{u\},\varnothing, N^{\rho})$ is a robust (up to $\rho$) controller for
%${\cal H}$.
%
%\end{example}

\section{Single-input Buck DC-DC Converter}\label{single-buck.tex}

\begin{figure}%[hbt!]
\begin{center}

\scalebox{0.45}{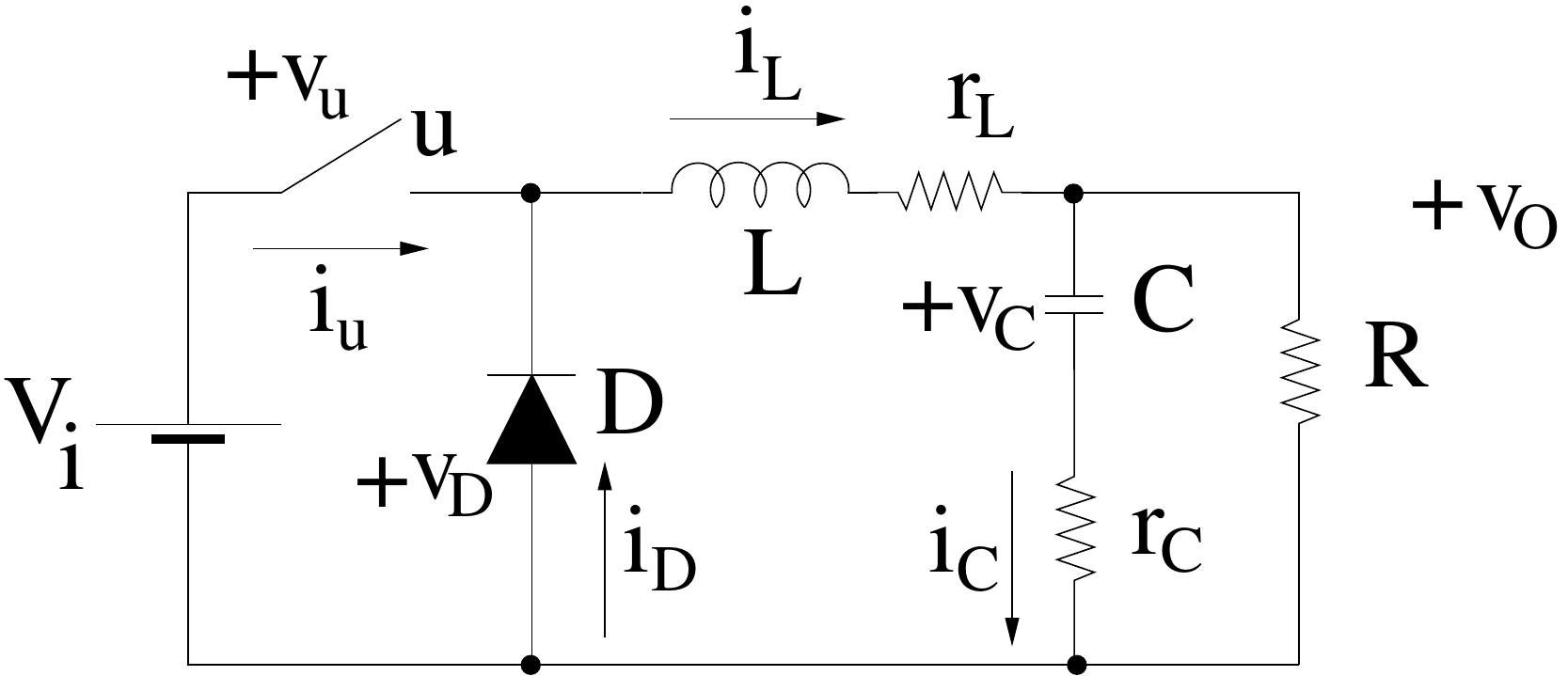}
\caption{Single-input buck DC-DC converter}\label{buck.eps}

\end{center}
\end{figure}

The buck DC-DC converter (Fig. \ref{buck.eps}) is a mixed-mode analog circuit
converting the DC input voltage ($V_{in}$ in Fig. \ref{buck.eps}) to a desired DC
output voltage ($v_O$ in Fig. \ref{buck.eps}). As an example, buck DC-DC
converters are used off-chip to scale down the typical laptop battery voltage
(12-24) to the just few volts needed by the laptop processor (e.g.
\cite{fuzzy-dc-dc-1996}) as well as on-chip to support \emph{Dynamic Voltage and
Frequency Scaling} (DVFS) in multicore processors (e.g.
\cite{gigascale-integration-07,buck-dc-dc-at-intel-pesc2004}). Because of its
widespread use, control schemas for buck DC-DC converters have been widely
studied (e.g. see
\cite{gigascale-integration-07,buck-dc-dc-at-intel-pesc2004,fuzzy-dc-dc-1996,time-optimal-dc-dc-2008}).
The typical software based approach (e.g. see \cite{fuzzy-dc-dc-1996}) is to
control the switch $u$ in Fig. \ref{buck.eps} (typically implemented with a
MOSFET) with a microcontroller.

Designing the software to run on the microcontroller to properly actuate the
switch is the control software design problem for the buck DC-DC converter in
our context.

The circuit in Fig.~\ref{buck.eps} can be modeled as a DTLHS ${\cal H}$ = ($X$,
$U$, $Y$, $N$). The circuit state variables are $i_L$ and $v_C$.  However we can
also use the pair $i_L$, $v_O$ as state variables in ${\cal H}$ model since
there is a linear relationship between $i_L$, $v_C$ and $v_O$, namely: $v_O \; =
\; \frac{r_C R}{r_C + R} i_L + \frac{R}{r_C + R} v_C$. Such considerations lead
to use the following sets of variables to model ${\cal H}$: $X$ $=$ $X^{r}$ $=$ 
$[i_L$, $v_O]$, $U$ $=$ $U^{b}$ $=$ $[u]$, $Y$ $=$ $Y^{r}\cup Y^{b}$ with
$Y^{r}$ $=$  $[i_u$, $v_u$, $i_D$, $v_D]$ and $Y^{b}$ $=$  $[q]$.  Note how
${\cal H}$ auxiliary variables $Y$ stem from the constitutive equations of the
switching elements (i.e. the switch $u$ and the diode D in Fig.~\ref{buck.eps}).
From a simple circuit analysis (e.g. see \cite{type2-fuzzy-dc-dc-2005}) we have
the following equations:

\begin{eqnarray}
\dot{i_L} & = & a_{1,1}i_L + a_{1,2} v_{O} + a_{1,3}v_D \\
\dot{v_{O}} & = & a_{2,1}i_L + a_{2,2}v_{O} + a_{2,3}v_D 
\end{eqnarray}

\noindent where the coefficients $a_{i, j}$ depend on the circuit parameters
$R$, $r_L$, $r_C$, $L$ and $C$ in the following way: $a_{1,1} = -\frac{r_L}{L}$,
$a_{1,2} = -\frac{1}{L}$, $a_{1,3} = -\frac{1}{L}$, $a_{2,1} = \frac{R}{r_c +
R}[-\frac{r_c r_L}{L} + \frac{1}{C}]$, $a_{2,2} = \frac{-1}{r_c + R}[\frac{r_c
R}{L} + \frac{1}{C}]$, $a_{2,3} = -\frac{1}{L}\frac{r_c R}{r_c + R}$. Using a
discrete time model with sampling time $T$ (writing $x'$ for $x(t+1)$) we have:

\begin{eqnarray}
{i_L}' & = & (1 + Ta_{1,1})i_L + Ta_{1,2}v_O + Ta_{1,3}v_D  \label{next-il.eq} \\
{v_O}' & = & Ta_{2,1}i_L + (1 + Ta_{2,2})v_O + Ta_{2,3}v_D. \label{next-vc.eq}
\end{eqnarray}

The algebraic constraints stemming from the constitutive equations of the
switching elements are the following:

\begin{center}
\begin{tabular}{lr}

\begin{minipage}{0.40\textwidth}
	\small
	\begin{eqnarray}
	q & \rightarrow & v_D = R_{\rm on}i_D \\
	q & \rightarrow & i_D \geq 0\\
	u & \rightarrow & v_u = R_{\rm on} i_u \\
	v_D & = & v_u - V_{in} \label{Vi.eq}
	\end{eqnarray}
\end{minipage} &

\begin{minipage}{0.45\textwidth}
	\small
	\begin{eqnarray}
	\bar{q} & \rightarrow &  v_D = R_{\rm off}i_D \\
	\bar{q} & \rightarrow & v_D \leq 0 \\
	\bar{u} & \rightarrow & v_u = R_{\rm off} i_u \label{mosfet-off-eq.eq}\\
	i_D & = & i_L - i_u \label{single.last-eq.eq}
	\end{eqnarray}
\end{minipage} \\   

\end{tabular}
\end{center}

The transition relation $N$ of ${\cal H}$ is given by the conjunction of the
constraints in Eqs.~\eqref{next-il.eq}--\eqref{single.last-eq.eq} and the
following explicit (safety) bounds: $-4 \leq i_L \leq 4 \land -1 \leq v_O \leq 7
\land -10^3 \leq i_D \leq 10^3 \land -10^3 \leq i_u \leq 10^3 \land -10^7 \leq
v_u \leq 10^7 \land -10^7 \leq v_D \leq 10^7$.

\subsection{Modelling Robustness on Input $V_{in}$ and Load
$R$}\label{robust.single.subsec}

In this section we address the problem of refining the model given in
Sect.~\ref{single-buck.tex} so as to require a controller for our single-input
buck to be robust to foreseen variations in the load $R$ and  in the power
supply $V_{in}$. That is, given tolerances $\rho_R$ and $\rho_{V_{in}}$, we
want the controller output by \qks \ for our single-input buck to work for any
$R \in [\max\{0, R(1 - \rho_R)\}, R(1 + \rho_R)]$ and any ${V_{in}} \in
[\max\{0, {V_{in}}(1 - \rho_{V_{in}})\}, {V_{in}}(1 + \rho_{V_{in}})]$.

Variations in the power supply are modeled  by replacing Eq.~\eqref{Vi.eq} in
Sect.~\ref{single-buck.tex} with the following:

\begin{center}
\begin{tabular}{lr}

\begin{minipage}{0.40\textwidth}
	\small
	\begin{eqnarray}
	v_D  \leq  v_u - V_{in}(1 - \rho_{V_{in}})
	\end{eqnarray}
\end{minipage} &

\begin{minipage}{0.45\textwidth}
	\small
	\begin{eqnarray}
	v_D  \geq  v_u - V_{in}(1 + \rho_{V_{in}})
	\end{eqnarray}
\end{minipage} \\   

\end{tabular}
\end{center}

Along the same lines, we may model also variations in the load $R$. However,
since $N$ dynamics is not linear in $R$,  much more work is needed (along the
lines of \cite{hypertech-hscc00}). To this aim, we proceed as follows.

The only equation depending on $R$ is Eq.~\eqref{next-vc.eq} of
Sect.~\ref{single-buck.tex}. Consider constants $a_{2,1}(R) = \frac{R}{r_c +
R}[-\frac{r_c r_L}{L} + \frac{1}{C}]$, $a_{2,2}(R) = \frac{-1}{r_c + R}[\frac{r_c
R}{L} + \frac{1}{C}]$, $a_{2,3}(R) = -\frac{1}{L}\frac{r_c R}{r_c + R}$
as (nonlinear) functions of $R$. It is easy to see that
$a_{2, 1}(R)$, $a_{2, 2}(R)$ are monotonically increasing functions for $R \in
\R^+$, while $a_{2, 3}(R)$ is monotonically decreasing for $R \in \R^+$. Thus,
if signs of $i_L, v_O, v_D$ are known, it is possible to replace
Eq.~\eqref{next-vc.eq} with two inequalities $v_O \geq Ta_{2,1}(R_{i_L}^{-})i_L
+ (1 + Ta_{2,2}(R_{v_O}^{-}))v_O + Ta_{2,3}(R_{v_D}^{-})v_D$ and $v_O \leq
Ta_{2,1}(R_{i_L}^{+})i_L + (1 + Ta_{2,2}(R_{v_O}^{+}))v_O +
Ta_{2,3}(R_{v_D}^{+})v_D$, being 

\begin{itemize}

	\item $R_w^{-} = $ {\bf if} $w \geq 0$ {\bf then}
$R(1 - \rho_R)$ {\bf else} $R(1 + \rho_R)$ and $R_w^{+} = $ {\bf if} $w \geq 0$ {\bf then}
$R(1 + \rho_R)$ {\bf else} $R(1 - \rho_R)$ for $w \in \{i_L, v_O\}$;

	\item $R_{v_D}^{-} = $ {\bf if} $v_D \geq 0$ {\bf then} $R(1 + \rho_R)$ {\bf else}
$R(1 - \rho_R)$ and $R_{v_D}^{+} = $ {\bf if} $v_D \geq 0$ {\bf then} $R(1 - \rho_R)$ {\bf else}
$R(1 + \rho_R)$.

\end{itemize}

%$d/dR[a_{2,1}(R)] = d/dR[\frac{R}{r_c + R}[-\frac{r_c r_L}{L} + \frac{1}{C}]] =
%\frac{[-\frac{r_c r_L}{L} + \frac{1}{C}]r_c}{(r_c + R)^2}$, $d/dR[a_{2,1}(R)] >
%0$ since $r_C > 0$ and $\frac{r_c r_L}{L} < \frac{1}{C}$ (because $\frac{L}{r_c
%r_L} > C$ since $2\times 10^{-2} > 5 \times 10^{-5}$).

%$d/dR[a_{2,2}(R)]$ $=$ $d/dR[\frac{-1}{r_c + R}[\frac{r_c R}{L} + \frac{1}{C}]]
%= d/dR[\frac{-r_c R}{Lr_c + LR} + \frac{-1}{Cr_c + CR}]$ $=$ \\$\frac{-r_cL(r_c
%+ R) + r_cRL}{L^2(r_c + R)^2} + \frac{C}{C^2(r_c + R)^2}$ $=$
%$\frac{-r_c^2L}{L^2(r_c + R)^2} + \frac{C}{C^2(r_c + R)^2}$ $=$
%$\frac{-r_c^2LC^2 + CL^2}{C^2L^2(r_c + R)^2}$, then $-r_c^2LC^2 + CL^2$
%evaluates to $ - 10^{-2} \times 2 \times 10^{-4} \times 25 \times 10^{-10} + 5
%\times 10^{-5} \times 4 \times 10^{-8} = -5 \times 10^{-15} + 2 \times 10^{-12}
%> 0$.

%$d/dR[a_{2,3}(R)] = d/dR[-\frac{1}{L}\frac{r_c R}{r_c + R}] = d/dR[-\frac{r_c
%R}{Lr_c + LR}] = \frac{-r_cL(r_c + R) + r_cRL}{L^2(r_c + R)^2} =
%\frac{-r_c^2}{L(r_c + R)^2} < 0$ 

This leads us to replace Eq.~\eqref{next-vc.eq} of Sect.~\ref{single-buck.tex}
with the equations in Fig.~\ref{buck-robust-model.fig}.

\begin{figure*}
\framebox[1.0\hsize]{
\begin{minipage}{1.0\hsize}
\begin{center}
\begin{tabular}{lr}

\begin{minipage}{0.50\textwidth}
	\small
	\begin{eqnarray}
	z_{i_L} & \to & i_L \geq 0\label{sign_il_pos.eq}\\
	z_{v_O} & \to & v_O \geq 0\label{sign_vo_pos.eq}\\
	z_{v_D} & \to & v_D \geq 0\label{sign_vd_pos.eq}
	\end{eqnarray}
\end{minipage} &

\begin{minipage}{0.45\textwidth}
	\small
	\begin{eqnarray}
	\overline{z_{i_L}} & \to & i_L \leq 0\label{sign_il_neg.eq}\\
	\overline{z_{v_O}} & \to & v_O \leq 0\label{sign_vo_neg.eq}\\
	\overline{z_{v_D}} & \to & v_D \leq 0\label{sign_vd_neg.eq}
	\end{eqnarray}
\end{minipage} \\   

\end{tabular}
\end{center}

\vspace*{-1cm}

\begin{eqnarray}
% z_{i_L} & \to & i_L \geq 0\label{sign_il_pos.eq}\\
% \overline{z_{i_L}} & \to & i_L \leq 0\label{sign_il_neg.eq}\\
% z_{v_O} & \to & v_O \geq 0\label{sign_vo_pos.eq}\\
% \overline{z_{v_O}} & \to & v_O \leq 0\label{sign_vo_neg.eq}\\
% z_{v_D} & \to & v_D \geq 0\label{sign_vd_pos.eq}\\
% \overline{z_{v_D}} & \to & v_D \leq 0\label{sign_vd_neg.eq}\\
 \overline{z_{ppp}} & \to & 1 - z_{i_L} + 1 - z_{v_O} + 1 - z_{v_D} \geq 1\label{comb_sign_0.eq}\\
 \overline{z_{pnp}} & \to & 1 - z_{i_L} + z_{v_O} + 1 - z_{v_D} \geq 1\label{comb_sign_1.eq}\\
 \overline{z_{ppn}} & \to & 1 - z_{i_L} + 1 - z_{v_O} + z_{v_D} \geq 1\label{comb_sign_2.eq}\\
 \overline{z_{pnn}} & \to & 1 - z_{i_L} + z_{v_O} + z_{v_D} \geq 1\label{comb_sign_3.eq}\\
 \overline{z_{npp}} & \to & z_{i_L} + 1 - z_{v_O} + 1 - z_{v_D} \geq 1\label{comb_sign_4.eq}\\
 \overline{z_{nnp}} & \to & z_{i_L} + z_{v_O} + 1 - z_{v_D} \geq 1\label{comb_sign_5.eq}\\
 \overline{z_{npn}} & \to & z_{i_L} + 1 - z_{v_O} + z_{v_D} \geq 1\label{comb_sign_6.eq}\\
 \overline{z_{nnn}} & \to & z_{i_L} + z_{v_O} + z_{v_D} \geq 1\label{comb_sign_7.eq}\\
% i_L' & = & (Ta_{1,1} + 1) i_L + Ta_{1,2} v_O + Tb_{1,1} v_D\label{il.eq}\\
 z_{ppp} & \to & v_O' \leq Ta_{2,1}^{(M)} i_L + (Ta_{2,2}^{(M)} + 1) v_O + Ta_{2,3}^{(m)} v_D\label{vo_sign_0_leq.eq}\\
 z_{ppp} & \to & v_O' \geq Ta_{2,1}^{(m)} i_L + (Ta_{2,2}^{(m)} + 1) v_O + Ta_{2,3}^{(M)} v_D\label{vo_sign_0_geq.eq}\\
 z_{ppn} & \to & v_O' \leq Ta_{2,1}^{(M)} i_L + (Ta_{2,2}^{(M)} + 1) v_O + Ta_{2,3}^{(M)} v_D\label{vo_sign_1_leq.eq}\\
 z_{ppn} & \to & v_O' \geq Ta_{2,1}^{(m)} i_L + (Ta_{2,2}^{(m)} + 1) v_O + Ta_{2,3}^{(m)} v_D\label{vo_sign_1_geq.eq}\\
 z_{pnp} & \to & v_O' \leq Ta_{2,1}^{(M)} i_L + (Ta_{2,2}^{(m)} + 1) v_O + Ta_{2,3}^{(m)} v_D\label{vo_sign_2_leq.eq}\\
 z_{pnp} & \to & v_O' \geq Ta_{2,1}^{(m)} i_L + (Ta_{2,2}^{(M)} + 1) v_O + Ta_{2,3}^{(M)} v_D\label{vo_sign_2_geq.eq}\\
 z_{pnn} & \to & v_O' \leq Ta_{2,1}^{(M)} i_L + (Ta_{2,2}^{(m)} + 1) v_O + Ta_{2,3}^{(M)} v_D\label{vo_sign_3_leq.eq}\\
 z_{pnn} & \to & v_O' \geq Ta_{2,1}^{(m)} i_L + (Ta_{2,2}^{(M)} + 1) v_O + Ta_{2,3}^{(m)} v_D\label{vo_sign_3_geq.eq}\\
 z_{npp} & \to & v_O' \leq Ta_{2,1}^{(m)} i_L + (Ta_{2,2}^{(M)} + 1) v_O + Ta_{2,3}^{(m)} v_D\label{vo_sign_4_leq.eq}\\
 z_{npp} & \to & v_O' \geq Ta_{2,1}^{(M)} i_L + (Ta_{2,2}^{(m)} + 1) v_O + Ta_{2,3}^{(M)} v_D\label{vo_sign_4_geq.eq}\\
 z_{npn} & \to & v_O' \leq Ta_{2,1}^{(m)} i_L + (Ta_{2,2}^{(M)} + 1) v_O + Ta_{2,3}^{(M)} v_D\label{vo_sign_5_leq.eq}\\
 z_{npn} & \to & v_O' \geq Ta_{2,1}^{(M)} i_L + (Ta_{2,2}^{(m)} + 1) v_O + Ta_{2,3}^{(m)} v_D\label{vo_sign_5_geq.eq}\\
 z_{nnp} & \to & v_O' \leq Ta_{2,1}^{(m)} i_L + (Ta_{2,2}^{(m)} + 1) v_O + Ta_{2,3}^{(m)} v_D\label{vo_sign_6_leq.eq}\\
 z_{nnp} & \to & v_O' \geq Ta_{2,1}^{(M)} i_L + (Ta_{2,2}^{(M)} + 1) v_O + Ta_{2,3}^{(M)} v_D\label{vo_sign_6_geq.eq}\\
 z_{nnn} & \to & v_O' \leq Ta_{2,1}^{(m)} i_L + (Ta_{2,2}^{(m)} + 1) v_O + Ta_{2,3}^{(M)} v_D\label{vo_sign_7_leq.eq}\\
 z_{nnn} & \to & v_O' \geq Ta_{2,1}^{(M)} i_L + (Ta_{2,2}^{(M)} + 1) v_O + Ta_{2,3}^{(m)} v_D\label{vo_sign_7_geq.eq}
% v_D & \leq & v_u - V_{in}(1 - \rho_{V_{in}})\label{vi_1.eq}\\
% v_D & \geq & v_u - V_{in}(1 + \rho_{V_{in}})\label{vi_2.eq}\\
% i_D & = & i_L - i_u\label{old_1.eq}\\
% q & \to & v_D = 0\label{old_2.eq}\\
% q & \to & i_D \geq 0\label{old_3.eq}\\
% \overline{q} & \to & v_D \leq 0\label{old_4.eq}\\
% \overline{q} & \to & v_D = R_{\rm off} i_D\label{old_5.eq}\\
% u & \to & v_u = 0\label{old_6.eq}\\
% \overline{u} & \to & v_u = R_{\rm off} i_u\label{old_7.eq}
\end{eqnarray}
\end{minipage}
}
\caption{DTLHS Buck Model Robust on $R$}\label{buck-robust-model.fig}
\end{figure*}

Note that, w.r.t. the model in Sect.~\ref{single-buck.tex}, in
Fig.~\ref{buck-robust-model.fig} we add to $Y^b$ 11 auxiliary boolean variables
$z_{i_L}$, $z_{v_O}$, $z_{v_D}$,  $z_{ppp}$, $ z_{ppp}$, $ z_{ppn}$, $ z_{ppn}$,
$ z_{pnp}$, $ z_{pnp}$, $z_{pnn}$, $ z_{pnn}$, $ z_{npp}$, $ z_{npp}$, $
z_{npn}$, $ z_{npn}$, $z_{nnp}$, $ z_{nnp}$, $ z_{nnn}$, $ z_{nnn}$ with the
following meaning. The boolean variable $z_{i_L}$ [$z_{v_O}$, $z_{v_D}$] is true
iff $i_L$ [${v_O}$, ${v_D}$] is positive (see Eqs.~(\ref{sign_il_pos.eq})
and~(\ref{sign_il_neg.eq}) [Eqs.~(\ref{sign_vo_pos.eq})
and~(\ref{sign_vo_neg.eq}), Eqs.~(\ref{sign_vd_pos.eq})
and~(\ref{sign_vd_neg.eq})]). The boolean variable $z_{abc}$, with $a, b, c \in
\{p, n\}$, is true iff  $(${\bf if} $a = p$ {\bf then} $i_L \geq 0$ {\bf else}
$i_L \leq 0) \land (${\bf if} $b = p$ {\bf then} $v_O \geq 0$ {\bf else} $v_O
\leq 0) \land (${\bf if} $c = p$ {\bf then} $v_D \geq 0$ {\bf else} $v_D \leq
0)$. This is stated by Eqs.~(\ref{comb_sign_0.eq})--(\ref{comb_sign_7.eq}).
Finally, we use boolean variables $z_{abc}$ as guards for the inequalities
replacing Eq.~\eqref{next-vc.eq} as stated before. This is done in
Eqs.~\eqref{vo_sign_0_leq.eq}--\eqref{vo_sign_7_geq.eq}.

%In Fig.~\ref{buck-robust-model.fig}, using $R^{(M)} = R(1 + \rho_R)$ and $R^{(m)} =
%R(1 - \rho_R)$, we define $a_{2,1}^{(M)} = \frac{R^{(M)}}{r_c +
%R^{(M)}}[-\frac{r_c r_L}{L} + \frac{1}{C}]$, $a_{2,2}^{(M)} = \frac{-1}{r_c +
%R^{(M)}}[\frac{r_c R^{(M)}}{L} + \frac{1}{C}]$,  $b_{2,1}^{(M)} =
%-\frac{1}{L}\frac{r_c R^{(M)}}{r_c + R^{(M)}}$, $a_{2,1}^{(m)} =
%\frac{R^{(m)}}{r_c + R^{(m)}}[-\frac{r_c r_L}{L} + \frac{1}{C}]$, $a_{2,2}^{(m)}
%= \frac{-1}{r_c + R^{(m)}}[\frac{r_c R^{(m)}}{L} + \frac{1}{C}]$, 
%$b_{2,1}^{(m)} = -\frac{1}{L}\frac{r_c R^{(m)}}{r_c + R^{(m)}}$.

\section{Multi-input Buck DC-DC Converter}\label{multi-buck.tex}

\begin{figure}%[hbt!]
\begin{center}

\scalebox{0.24}{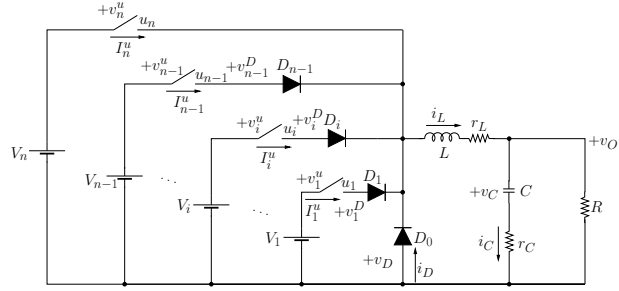}
\caption{Multi-input Buck DC-DC converter}\label{multibuck.eps}

\end{center}
\end{figure}

A multi-input buck DC-DC converter~\cite{multin-buck-dcdc-2010}
(Fig.~\ref{multibuck.eps}),  consists of $n$ power supplies with voltage values
$V_1 < \ldots < V_n$, $n$ switches with voltage values $v_1^u, \ldots, v_n^u$
and current values $I_1^{u}, \ldots, I_n^{u}$, and $n$ input diodes $D_0,
\ldots, D_{n - 1}$ with voltage values $v_0^D, \ldots, v_{n - 1}^D$ and current
values $i_0^D, \ldots, i_{n - 1}^D$ (in the following, we will also write $v_D$
for $v_0^D$ and $i_D$ for $i_0^D$). As for the converter in
Sect.~\ref{single-buck.tex}, the state variables are $i_L$ and $v_O$.
Differently from the converter in Sect.~\ref{single-buck.tex}, the action
variables are $u_1, \ldots, u_n$, thus a control software for the $n$-input buck
dc-dc converter has to properly actuate the switches $u_1, \ldots, u_n$.
%Constant values are the same given in Sect.~\ref{single-buck.tex}.

We model our $n$-input buck DC-DC converter with DTLHS ${\cal H}$ = ($X,$ $U,$
$Y,$ $N$), where $X = X^r = [i_L,$ $v_O],$ $U = U^b = [u_1,$ $\ldots,$ $u_n],$
and $Y = Y^r \cup Y^b$ with $Y^r = [v_D,$ $v_1^D,$ $\ldots, v_{n - 1}^{D},$
$i_D,$ $I_1^{u},$ $\ldots,$ $I_n^{u},$ $ v_1^u,$ $\ldots,$ $v_n^u]$ and $Y^b =
[q_0,$ $\ldots,$ $q_{n - 1}]$. As for the predicate $N$, from a simple circuit
analysis (e.g. see~\cite{type2-fuzzy-dc-dc-2005}) we have that state variables
constraints are the same as Eqs.~\eqref{next-il.eq} and~\eqref{next-vc.eq} of
the converter in Sect.~\ref{single-buck.tex}. 

The algebraic constraints stemming from the constitutive equations of the
switching elements are the following (where $i$ and $j$ range in $[n - 1]$ and
$[n]$ respectively):
%In addition
%to Eqs.~\eqref{next-il.eq} and~\eqref{next-vc.eq} of Sect.~\ref{single-buck.tex}, the
%following Eqs.~\eqref{mbuck-begin.eq}--\eqref{mbuck-end.eq} must hold. 

\begin{center}
\begin{tabular}{lr}

\begin{minipage}{0.45\textwidth}
\small
\begin{eqnarray}
q_0  & \rightarrow & v_D = R_{\rm on} i_D \label{mbuck-begin.eq} \\
q_0  & \rightarrow & i_D \geq 0 \\
q_i & \rightarrow & v_i^D = R_{\rm on} I_i^u \\
q_i & \rightarrow & I_i^u \geq 0 \\
u_j & \rightarrow & v_j^u = R_{\rm on} I_j^u\\
i_L &  = & i_D + \sum_{i=1}^{n}I_i^{u}
\end{eqnarray}
\end{minipage} &

\begin{minipage}{0.45\textwidth}
\small
\begin{eqnarray}
\bar{q}_0 & \rightarrow &  v_D = R_{\rm off}i_D \\
\bar{q}_0 & \rightarrow & v_D \leq 0 \\
\bar{q}_i & \rightarrow &  v_i^D = R_{\rm off}I_i^u \\
\bar{q}_i & \rightarrow & v_i^D \leq 0 \\
\bar{u}_j & \rightarrow & v_j^u = R_{\rm off} I_j^u\\
v_D  & = & v_i^u + v_{i}^{D} - V_i \label{Vi.i.eq} \\
v_D &  = & v_n^u - V_n\label{mbuck-end.eq}\label{Vn.eq}
\end{eqnarray}
\end{minipage} \\

\end{tabular}
\end{center}

Finally, $N$ is given by the conjunction of Eqs.~\eqref{next-il.eq}
and~\eqref{next-vc.eq} of Sect.~\ref{single-buck.tex},
Eqs.~\eqref{mbuck-begin.eq}--\eqref{mbuck-end.eq} and the following explicit
(safety) bounds: $-4 \leq i_L \leq 4 \land -1 \leq v_O \leq 7 \land -10^3 \leq
i_D \leq 10^3 \land \;\bigwedge_{i = 1}^n -10^3 \leq I_i^u \leq 10^3 \land
\;\bigwedge_{i = 1}^n -10^7 \leq v_i^u \leq 10^7 \land \;\bigwedge_{i = 0}^{n -
1} -10^7$ $\leq v_i^D \leq 10^7$.

\subsection{Modelling Robustness on Inputs $V_{i}$ and Load
$R$}\label{robust.multi.subsec}

In this section we address the problem of refining the model given in
Sect.~\ref{multi-buck.tex} so as to require a controller for our multi-input
buck to be robust to foreseen variations in the load $R$ and  in the power
supplies $V_i$ (for $i \in [n]$). As it is explained in
Sect.~\ref{robust.single.subsec}, given tolerances $\rho_R$ and $\rho_{V_i}$
(for $i \in [n]$), we want the controller output by \qks \ for our multi-input
buck to work for any $R \in [\max\{0, R(1 - \rho_R)\}, R(1 + \rho_R)]$ and any
${V_i} \in [\max\{0, {V_i}(1 - \rho_{V_i})\}, {V_i}(1 + \rho_{V_i})]$ (for $i
\in [n]$).

Variations in the power supplies are modeled  by replacing Eqs.~\eqref{Vi.i.eq}
and~\eqref{Vn.eq} in Sect.~\ref{multi-buck.tex} with the following (where $i$
ranges in $[n - 1]$):

\begin{center}
\begin{tabular}{lr}

\begin{minipage}{0.50\textwidth}
	\small
	\begin{eqnarray}
	v_D  \leq v_i^u + v_{i}^{D} - V_i(1 - \rho_{V_{i}})\\
	v_D  \geq v_i^u + v_{i}^{D} - V_i(1 + \rho_{V_{i}})
	\end{eqnarray}
\end{minipage} &

\begin{minipage}{0.45\textwidth}
	\small
	\begin{eqnarray}
	v_D \leq v_n^u - V_n(1 - \rho_{V_{n}})\\
	v_D \geq v_n^u - V_n(1 + \rho_{V_{n}})
	\end{eqnarray}
\end{minipage} \\   

\end{tabular}
\end{center}

As for the robustness w.r.t. the load $R$, since the only equation depending on
$R$ is Eq.~\eqref{next-vc.eq} of Sect.~\ref{single-buck.tex}, which also holds
for the multi-input buck, the same reasoning of Sect.~\ref{robust.single.subsec}
may be applied. Thus, we have to replace Eq.~\eqref{next-vc.eq} of
Sect.~\ref{single-buck.tex} with the equations in
Fig.~\ref{buck-robust-model.fig}.

\section{Experimental Results}\label{expres.tex}

In this section we present our experimental results about running
\qks~\cite{qks-cav2010} on the buck models described in
Sects.~\ref{single-buck.tex} and~\ref{multi-buck.tex}. Namely, we will present
experimental results on the robust model for the single-input buck described in
Sect.~\ref{robust.single.subsec} (Sect.~\ref{robust.single.expres.subsec}) and
on the (non-robust) model for the multi-buck described in
Sect.~\ref{multi-buck.tex} (Sect.~\ref{multi.expres.subsec}). All experiments
run on an Intel 3.0 GHz hyperthreaded Quad Core Linux PC with 8 GB of RAM.

\subsection{Single-input Buck}\label{robust.single.expres.subsec}

We run \qks \ on the single-input buck model taking into account foreseen
variations in the load $R$ and  in the power supply $V_{in}$ (see
Sect.~\ref{robust.single.subsec}). Since \qks \ also require as input the number
of AD bits $b$ (see~\cite{qks-cav2010} for details), we run multiple times \qks
\ for different values of $b$, each time obtaining a controller $K^b$. All other
constants introduced in Sect.~\ref{single-buck.tex} are fixed as follows: $T =
10^{-6}$ secs,  $L = 2 \cdot 10^{-4}$ H, $r_L = 0.1$ ${\rm \Omega}$, $r_C = 0.1$
${\rm \Omega}$,  $R = 5$ ${\rm \Omega}$,  $C = 5 \cdot 10^{-5}$ F,  $V_i = 15$
V, $\rho_R = \rho_{V_{in}} = 25\%$, $R_{\rm on} = 0$ ${\rm \Omega}$, $R_{\rm
off} = 10^4$ ${\rm \Omega}$.

\begin{sidewaystable}
\centering
\small
\caption{Single-input buck DC-DC converter: control abstraction and controller synthesis results.}\label{robust.single.expres.tab.tex}
\begin{tabular}{ccccccccccc}
\toprule
& \multicolumn{5}{c}{Control Abstraction} & \multicolumn{2}{c}{Controller Synthesis} & \multicolumn{3}{c}{Total}\\
\cmidrule(r){2-6}\cmidrule(r){7-8}\cmidrule(r){9-11} $b$ & CPU & MEM & Arcs & MaxLoops & LoopFrac & CPU & $|K|$ & CPU & MEM & $\mu$\\
\midrule
8 & 1.95e+03 & 4.41e+07 & 6.87e+05 & 2.55e+04 & 0.00333 & 2.10e-01 & 1.39e+02 & 1.96e+03 & 4.46e+07 & {\sc Unk}\\
9 & 9.55e+03 & 5.67e+07 & 3.91e+06 & 1.87e+04 & 0.00440 & 2.64e+01 & 3.24e+03 & 9.58e+03 & 7.19e+07 & {\sc Sol}\\
10 & 1.42e+05 & 8.47e+07 & 2.61e+07 & 2.09e+04 & 0.00781 & 7.36e+01 & 1.05e+04 & 1.42e+05 & 1.06e+08 & {\sc Sol}\\
11 & 8.76e+05 & 1.11e+08 & 2.15e+08 & 2.26e+04 & 0.01435 & 2.94e+02 & 2.88e+04 & 8.76e+05 & 2.47e+08 & {\sc Sol}\\
\bottomrule
\end{tabular}
\end{sidewaystable}

\begin{table}
\centering
\small
\caption{Single-input buck DC-DC converter: number of MILPs and time to solve them}\label{robust.single.expres.milps.1.tab.tex}
\begin{tabular}{*{7}{c}}
\toprule
& \multicolumn{3}{c}{$b=8$} & \multicolumn{3}{c}{$b=9$}\\
\cmidrule(r){2-4}\cmidrule(r){5-7}
MILP & Num & Avg & Time & Num & Avg & Time \\
\midrule
1 & 6.6e+04 & 7.0e-05 & 4.6e+00 & 2.6e+05 & 7.0e-05 & 1.8e+01 \\
2 & 4.0e+05 & 1.5e-03 & 3.3e+02 & 1.6e+06 & 1.4e-03 & 1.1e+03 \\
3 & 2.3e+05 & 9.1e-04 & 2.1e+02 & 9.2e+05 & 9.2e-04 & 8.4e+02 \\
4 & 7.8e+05 & 9.9e-04 & 7.7e+02 & 4.4e+06 & 1.0e-03 & 4.5e+03 \\
5 & 4.3e+05 & 2.8e-04 & 1.2e+02 & 1.7e+06 & 2.8e-04 & 4.9e+02 \\
\bottomrule
\end{tabular}
\end{table}

\begin{table}
\centering
\small
\caption{Single-input buck DC-DC converter: number of MILPs and time to solve them (continuation of Tab.~\ref{robust.single.expres.milps.1.tab.tex})}\label{robust.single.expres.milps.2.tab.tex}
\begin{tabular}{*{7}{c}}
\toprule
& \multicolumn{3}{c}{$b=10$} & \multicolumn{3}{c}{$b=11$}\\
\cmidrule(r){2-4}\cmidrule(r){5-7}
MILP & Num & Avg & Time & Num & Avg & Time \\
\midrule
1 & 1.0e+06 & 2.7e-04 & 2.8e+02 & 4.2e+06 & 2.3e-04 & 9.7e+02\\
2 & 6.4e+06 & 3.8e-03 & 1.3e+04 & 2.5e+07 & 3.3e-03 & 4.6e+04\\
3 & 3.7e+06 & 3.0e-03 & 1.1e+04 & 1.5e+07 & 2.6e-03 & 3.8e+04\\
4 & 3.0e+07 & 2.6e-03 & 7.8e+04 & 2.6e+08 & 2.2e-03 & 5.7e+05\\
5 & 6.8e+06 & 1.8e-03 & 1.3e+04 & 2.7e+07 & 1.6e-03 & 4.2e+04\\
\bottomrule
\end{tabular}
\end{table}

Tabs.~\ref{robust.single.expres.tab.tex},~\ref{robust.single.expres.milps.1.tab.tex}
and~\ref{robust.single.expres.milps.2.tab.tex} show our experimental results.
Columns in Tab.~\ref{robust.single.expres.tab.tex} have the following meaning. 
Column $b$ shows the number of AD bits (see~\cite{qks-cav2010} for details). 
Columns labeled {\em Control Abstraction} show performance for control
abstraction computation (see~\cite{qks-cav2010} for details) and they show
running time (column {\em CPU}, in secs),  memory usage ({\em MEM}, in bytes), 
the number of transitions in the generated control abstraction ({\em Arcs}), the
number of self-loops in the maximum control abstraction ({\em MaxLoops}),  and
the fraction of loops that are kept in the minimum control abstraction  w.r.t.
the number of loops in the maximum control abstraction ({\em LoopFrac}).

Columns labeled {\em Controller Synthesis} show the computation time (column
{\em CPU}, in secs) for the generation of $K^b$, and the size of its OBDD
representation ({\em OBDD}, number of nodes). The latter is also  the size
(number of lines) of $K^b$ C code  synthesized implementation.  Finally, columns
labeled {\em Total} show the total computation time (column {\em CPU}, in secs) 
and the memory ({\em MEM}, in bytes)  for the whole process (i.e., control
abstraction plus controller source code generation), as well as the final
outcome $\mu\in\{${\sc Sol}, {\sc NoSol}, {\sc Unk}$\}$ of \qks \
(see~\cite{qks-cav2010} for details).

For each MILP problem solved in \qks \ (see~\cite{qks-cav2010} for details),
Tabs.~\ref{robust.single.expres.milps.1.tab.tex} 
and~\ref{robust.single.expres.milps.2.tab.tex} show (as a function of $b$) the
total and the average CPU time (in seconds) spent solving MILP problem
instances, together with the number of MILP instances solved. Columns in
Tabs.~\ref{robust.single.expres.milps.1.tab.tex} 
and~\ref{robust.single.expres.milps.2.tab.tex} have the following meaning: {\em
Num} is the number of times that the MILP problem of the given type is called,
{\em Time} is the total CPU time (in secs) needed to solve all the {\em Num}
instances of the MILP problem of the given type, and {\em Avg} is the average
CPU time (in secs), i.e. the ratio between columns {\em Time} and {\em Num}.
Each row in Tabs.~\ref{robust.single.expres.milps.1.tab.tex} 
and~\ref{robust.single.expres.milps.2.tab.tex} refer to a type of MILP problem
solved, see~\cite{qks-cav2010} for details.

\begin{figure}
  \centering
  \includegraphics[width=1\textwidth]{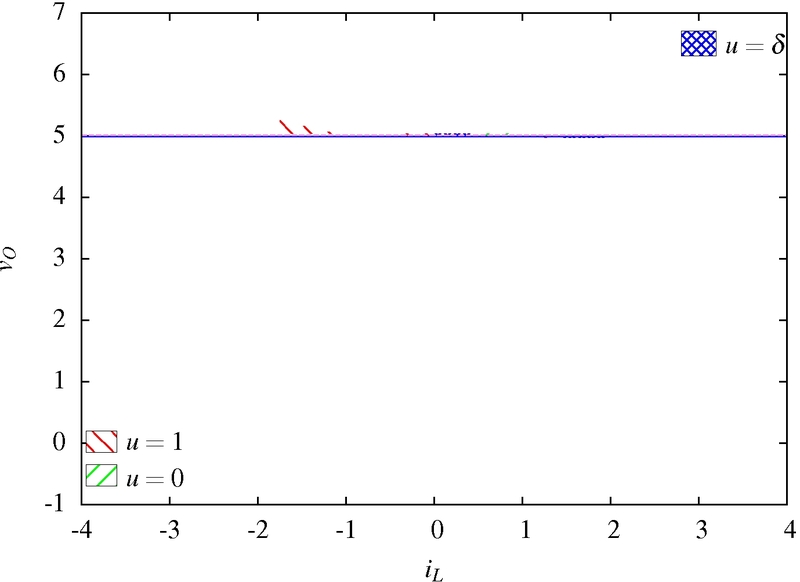}
  \caption{Single-input robust buck: controlled region with $b=8$ bits}
  \label{robust.single.ctrl-reg.8.eps}
\end{figure}

\begin{figure}
  \centering
  \includegraphics[width=1\textwidth]{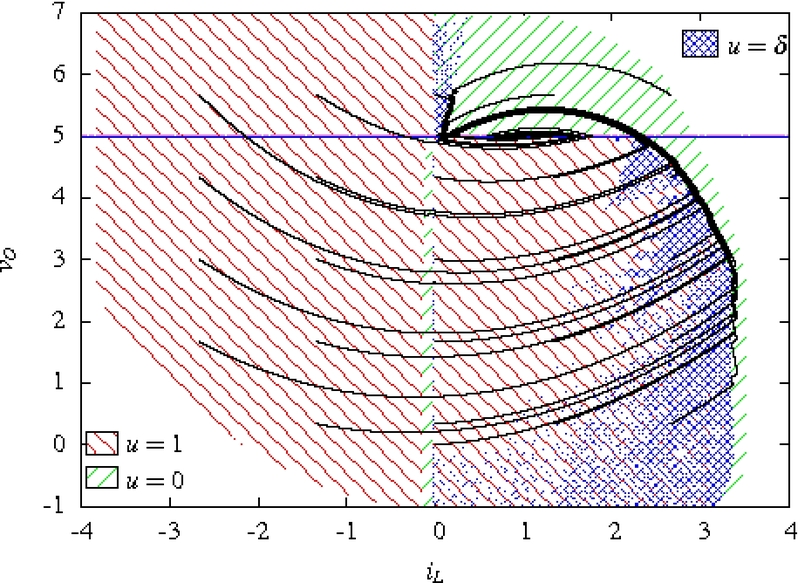}
  \caption{Single-input robust buck: controlled region with $b=9$ bits}
  \label{robust.single.ctrl-reg.9.eps}
\end{figure}

\begin{figure}
  \centering
  \includegraphics[width=1\textwidth]{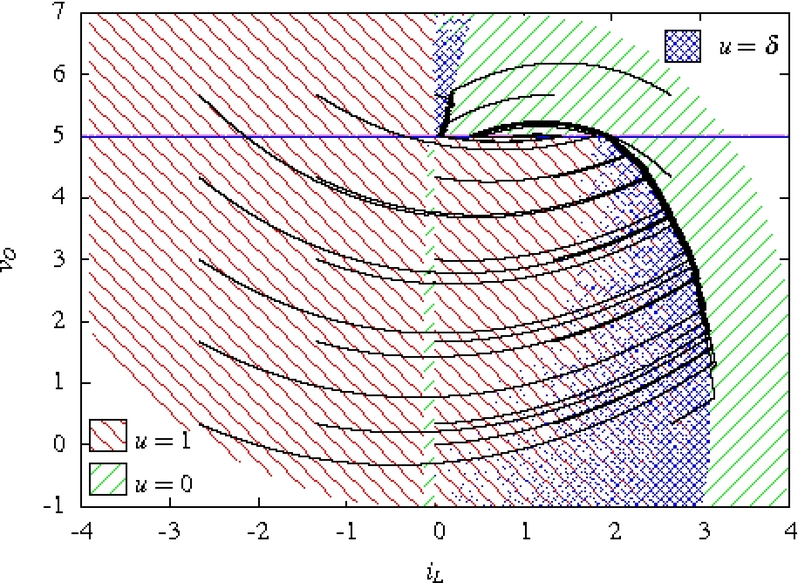}
  \caption{Single-input robust buck: controlled region with $b=10$ bits}
  \label{robust.single.ctrl-reg.10.eps}
\end{figure}

\begin{figure}
  \centering
  \includegraphics[width=1\textwidth]{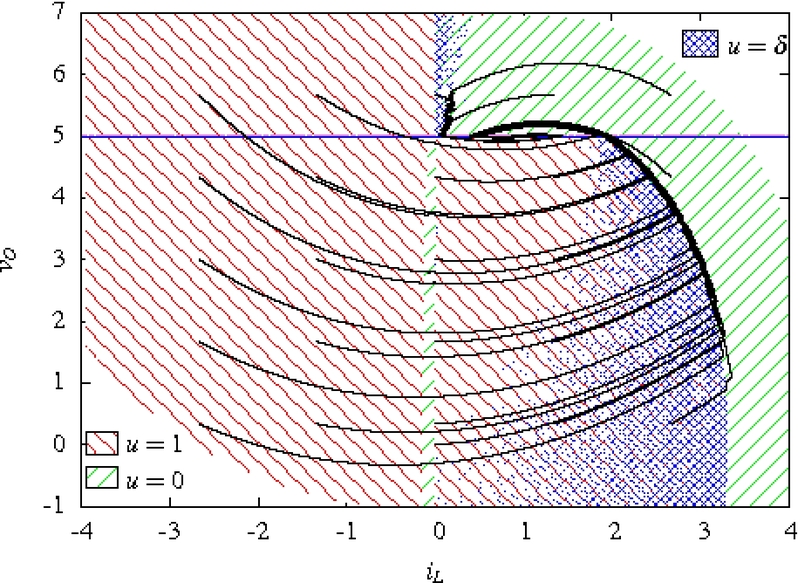}
  \caption{Single-input robust buck: controlled region with $b=11$ bits}
  \label{robust.single.ctrl-reg.11.eps}
\end{figure}

Finally, in
Figs.~\ref{robust.single.ctrl-reg.8.eps}--\ref{robust.single.ctrl-reg.11.eps} we
show the guaranteed operational range ({\em controlled regions},
see~\cite{qks-cav2010} for details) of the controllers generated for the
single-input buck by \qks.

\subsection{Multi-input Buck}\label{multi.expres.subsec}

We run \qks \ on the multi-input buck model described in 
Sect.~\ref{multi-buck.tex}. Differently from
Sect.~\ref{robust.single.expres.subsec}, we fix the number of AD bits $b$ for
\qks, namely $b = 10$. On the other hand, we run multiple times \qks \ by
varying the number $n$ of inputs for the multi-input buck. As for input voltages, we
have $V_i = 10i$ V for all $i \in [n]$. All other constants
introduced in Sect.~\ref{multi-buck.tex} are fixed as in
Sect.~\ref{robust.single.expres.subsec}.

\begin{sidewaystable}
\centering
\small
\caption{Multi-input buck DC-DC converter: control abstraction and controller synthesis results}\label{multi.expres.tab.tex}
\begin{tabular}{ccccccccccc}
\toprule
& \multicolumn{5}{c}{Control Abstraction} & \multicolumn{2}{c}{Controller Synthesis} & \multicolumn{3}{c}{Total}\\
\cmidrule(r){2-6}\cmidrule(r){7-8}\cmidrule(r){9-11} $n$ & CPU & MEM & Arcs & MaxLoops & NoLoopsPerc & CPU & $|K|$ & CPU & MEM & $\mu$\\
\midrule
1 & 2.88e+04 & 6.41e+07 & 7.38e+06 & 1.91e+04 & 0.00377 & 1.97e+01 & 1.21e+04 & 2.88e+04 & 8.35e+07 & {\sc Sol}\\
2 & 8.94e+04 & 7.63e+07 & 1.47e+07 & 1.91e+04 & 0.00743 & 2.66e+01 & 2.52e+04 & 8.94e+04 & 8.25e+07 & {\sc Sol}\\
3 & 2.46e+05 & 9.47e+07 & 2.93e+07 & 1.90e+04 & 0.01162 & 3.66e+01 & 3.47e+04 & 2.46e+05 & 1.05e+08 & {\sc Sol}\\
4 & 6.43e+05 & 9.51e+07 & 5.84e+07 & 1.88e+04 & 0.00330 & 5.32e+01 & 4.31e+04 & 6.43e+05 & 0.00e+00 & {\sc Sol}\\
\bottomrule
\end{tabular}
\end{sidewaystable}

\begin{table}
  \centering
  \small
  \caption{Multi-input buck DC-DC converter: number of MILPs and time to solve them}\label{multi.expres.milps.1.tab.tex}
  \begin{tabular}{*{7}{c}}
    \toprule
    & \multicolumn{3}{c}{$n=1$}& \multicolumn{3}{c}{$n=2$}\\
    \cmidrule(r){2-4}\cmidrule(r){5-7}
    MILP & Num & Avg & Time  & Num & Avg & Time \\
    \midrule
1 & 1.0e+06 & 2.0e-04 & 2.1e+02 & 1.0e+06 & 2.1e-04 & 2.2e+02\\

2 & 6.4e+06 & 1.4e-03 & 5.1e+03 & 1.3e+07 & 1.9e-03 & 1.6e+04\\

3 & 3.7e+06 & 8.8e-04 & 3.2e+03 & 7.4e+06 & 1.6e-03 & 1.1e+04\\

4 & 8.7e+06 & 1.0e-03 & 8.9e+03 & 1.7e+07 & 1.7e-03 & 2.8e+04\\

5 & 6.9e+06 & 6.8e-04 & 4.6e+03 & 1.4e+07 & 1.1e-03 & 1.5e+04\\

    \bottomrule
  \end{tabular}
\end{table}

\begin{table}
  \centering
  \small
  \caption{Multi-input buck DC-DC converter: number of MILPs and time to solve them (continuation of
  Tab.~\ref{multi.expres.milps.1.tab.tex})}\label{multi.expres.milps.2.tab.tex}
  \begin{tabular}{*{7}{c}}
    \toprule
    & \multicolumn{3}{c}{$n=3$}& \multicolumn{3}{c}{$n=4$}\\
    \cmidrule(r){2-4}\cmidrule(r){5-7}
    MILP & Num & Avg & Time  & Num & Avg & Time \\
    \midrule
1 & 1.0e+06 & 2.1e-04 & 2.2e+02 & 1.0e+06 & 2.2e-04 & 2.3e+02\\

2 & 2.5e+07 & 3.0e-03 & 4.6e+04 & 5.1e+07 & 4.5e-03 & 1.2e+05\\

3 & 1.5e+07 & 2.2e-03 & 3.2e+04 & 2.9e+07 & 2.9e-03 & 8.6e+04\\

4 & 3.2e+07 & 2.4e-03 & 7.9e+04 & 6.3e+07 & 3.2e-03 & 2.0e+05\\

5 & 2.7e+07 & 1.6e-03 & 4.3e+04 & 5.5e+07 & 2.1e-03 & 1.1e+05\\

    \bottomrule
  \end{tabular}
\end{table}

Tabs.~\ref{multi.expres.tab.tex},~\ref{multi.expres.milps.1.tab.tex}
and~\ref{multi.expres.milps.2.tab.tex} show our experimental results. Columns in
Tab.~\ref{multi.expres.tab.tex} have the following meaning.  Column $n$ shows
the number of inputs of the multi-input buck (see Sect.~\ref{multi-buck.tex} for
details).  All other columns of Tab.~\ref{multi.expres.tab.tex}, as well as of
Tabs.~\ref{multi.expres.milps.1.tab.tex} and~\ref{multi.expres.milps.2.tab.tex}
have the same meaning of the same columns of
Tabs.~\ref{robust.single.expres.tab.tex},~\ref{robust.single.expres.milps.1.tab.tex}
and~\ref{robust.single.expres.milps.2.tab.tex}.

\begin{figure}
  \centering
  \includegraphics[angle=-90,width=1\textwidth]{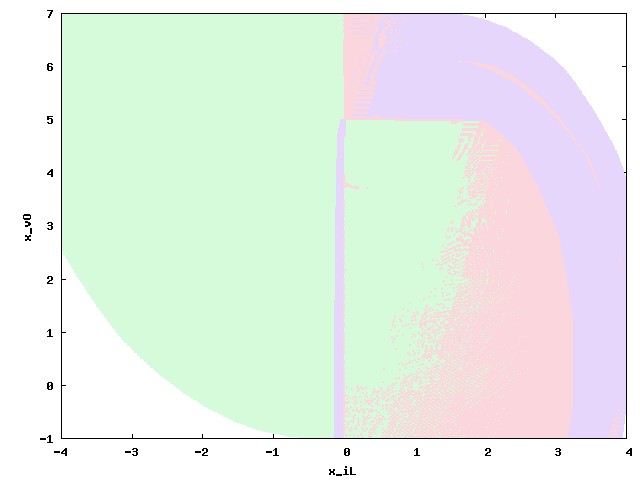}
  \caption{Multi-input buck: controlled region with $n=1$ inputs}
  \label{multi.ctrl-reg.1.eps}
\end{figure}

\begin{figure}
  \centering
  \includegraphics[angle=-90,width=1\textwidth]{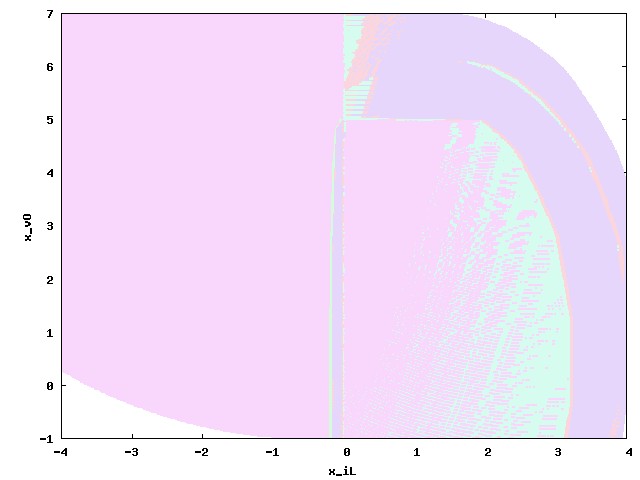}
  \caption{Multi-input buck: controlled region with $n=2$ inputs}
  \label{multi.ctrl-reg.2.eps}
\end{figure}

\begin{figure}
  \centering
  \includegraphics[angle=-90,width=1\textwidth]{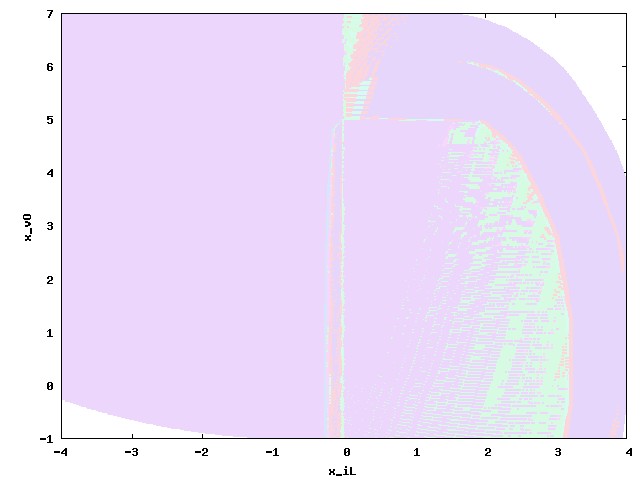}
  \caption{Multi-input buck: controlled region with $n=3$ inputs}
  \label{multi.ctrl-reg.3.eps}
\end{figure}

\begin{figure}
  \centering
  \includegraphics[angle=-90,width=1\textwidth]{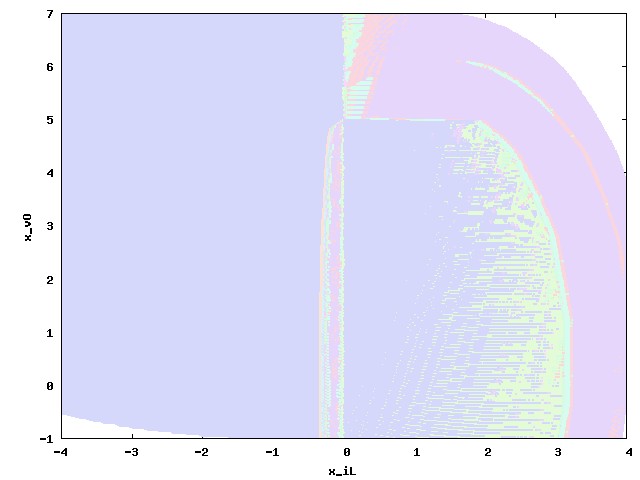}
  \caption{Multi-input buck: controlled region with $n=4$ inputs}
  \label{multi.ctrl-reg.4.eps}
\end{figure}

Finally, in Figs.~\ref{multi.ctrl-reg.1.eps}--\ref{multi.ctrl-reg.4.eps} we
show the guaranteed operational range ({\em controlled regions},
see~\cite{qks-cav2010} for details) of the controllers generated for the
multi-input buck by \qks.

\section{Conclusions}\label{conclu.tex}

We presented experimental results on using the \qks \ tool~\cite{qks-cav2010},
to support a \emph{Formal Model Based Design} approach to control software. 
Our experiments have been carried out on two versions of the buck DC-DC
converter, namely the single-input and the multi-input versions. We also showed
how robust controllers may be generated for such bucks, namely by taking into
account also foreseen variations on some important buck parameters such as load
and input power supplies.

\bibliographystyle{plain}
\bibliography{modelchecking}

\end{document}